\documentclass[aps,prb,twocolumn,superscriptaddress,floatfix,showkeys]{revtex4-2}
\usepackage{etoolbox}
\usepackage{graphics,graphicx} 
\usepackage{natbib}
\usepackage{subfig}
\usepackage{float}
\usepackage{placeins} 
\usepackage{ragged2e}  
\usepackage{caption}
\usepackage{xcolor}
\usepackage{amsmath}
\usepackage{amssymb}
\usepackage{braket}
\usepackage{blindtext}
\usepackage{bm}
\usepackage{subfig}  
\usepackage{xcolor}
\usepackage{algorithm}
\usepackage{algpseudocode}
\usepackage{hyperref}
\hypersetup{
    colorlinks=true,
    citecolor=blue, 
    linkcolor=red, 
    urlcolor= blue 
}

\usepackage{MnSymbol}

\usepackage[cal=boondox,scr=boondoxo]{mathalfa}

\begin{document}


\title{Longitudinal-Field-Driven Transition in a non-integrable Non-Hermitian Transverse-Field Ising Chain via RBMs}


\author{M.E. Ateuafack}
\email{esouamath@yahoo.fr}
\affiliation{Department of Electrical and Electronics Engineering, College of Technology, University of Buea, P.O. Box 63, Buea, Cameroon}
\affiliation{Promotion Centre of Research for Technological Advancement and Sustainable Development (PCR-TASD), P.O. Box 55, Maroua, Cameroon}
\affiliation{Quantum Materials and Computing Group (QMaCG), P.O. Box 70, Bambili, North West Region, Cameroon}

\author{M.R. Kamsap}
\affiliation{Laboratoire d'Optique, Laser et Applications, Unité de Recherche en Physique, Université des Sciences et Techniques de Masuku, B.P. 943, Franceville, Gabon}
\affiliation{Département de Physique, Faculté des Sciences, Université des Sciences et Techniques de Masuku, B.P. 943, Franceville, Gabon}

\author{J.T. Diffo}
\affiliation{Unité de Recherche de Matière Condensée, d'Électronique et de Traitement de Signal (URMACETS), Department of Physics, Faculty of Science, University of Dschang, P.O. Box 67, Dschang, Cameroon}
\affiliation{Department of Physics, Higher Teachers Training College, The University of Maroua, P.O. Box 55, Maroua, Cameroon}
\affiliation{Promotion Centre of Research for Technological Advancement and Sustainable Development (PCR-TASD), P.O. Box 55, Maroua, Cameroon}
\affiliation{Quantum Materials and Computing Group (QMaCG), P.O. Box 70, Bambili, North West Region, Cameroon}

\author{L.C. Fai}
\affiliation{Unité de Recherche de Matière Condensée, d'Électronique et de Traitement de Signal (URMACETS), Department of Physics, Faculty of Science, University of Dschang, P.O. Box 67, Dschang, Cameroon}
\affiliation{Quantum Materials and Computing Group (QMaCG), P.O. Box 70, Bambili, North West Region, Cameroon}

\author{M.N. Jipdi}
\affiliation{Unité de Recherche de Matière Condensée, d'Électronique et de Traitement de Signal (URMACETS), Department of Physics, Faculty of Science, University of Dschang, P.O. Box 67, Dschang, Cameroon}
\affiliation{Department of Physics, Higher Teachers Training College, University of Bamenda, P.O. Box 11, Bamenda, Cameroon}
\affiliation{Promotion Centre of Research for Technological Advancement and Sustainable Development (PCR-TASD), P.O. Box 55, Maroua, Cameroon}
\affiliation{Quantum Materials and Computing Group (QMaCG), P.O. Box 70, Bambili, North West Region, Cameroon}


\date{\today}

\begin{abstract}
We investigate the ground-state properties and quantum critical behavior of a non-Hermitian transverse-field Ising chain subjected to longitudinal and complex transverse magnetic fields. To address this interacting many-body problem, we employ real-valued neural quantum states based on Restricted Boltzmann Machines (RBMs), optimized using Variational Monte Carlo (VMC) sampling. Spectral analysis of finite chains reveals exceptional points associated with spontaneous parity-time (PT) symmetry breaking. A real-valued RBM framework is developed to reconstruct the ground-state eigenstates of the non-Hermitian Hamiltonian. Benchmark comparisons with exact diagonalization demonstrate that the RBM approach accurately reproduces the ground-state energy, magnetization, and spin-spin correlations. Extending the analysis to larger system sizes, we identify a non-Hermitian quantum phase transition characterized by PT-symmetry breaking and the emergence of magnetic order. Our results establish real-valued neural quantum states as an efficient and scalable framework for investigating critical phenomena in interacting non-Hermitian quantum systems.

\end{abstract}


\maketitle


\section{Introduction}

Understanding the behavior of interacting quantum many-body systems subjected to external fields remains one of the central problems in condensed matter physics\cite{Lieb1961,Pfeuty1970,Dutta2015,Sachdev2011}. Quantum spin chains, in particular, provide fundamental platforms for exploring quantum phase transitions, collective excitations, critical phenomena, and non-equilibrium dynamics. Among these models, the transverse-field Ising model (TFIM) occupies a prominent position because of its simplicity and its ability to capture essential features of quantum magnetism and criticality \cite{Dutta2015,Sachdev2011}.

The introduction of a longitudinal magnetic field fundamentally alters the physics of the transverse-field Ising model. While the conventional model remains exactly solvable through the Jordan-Wigner transformation, the longitudinal field destroys its free-fermion structure by generating interacting fermionic terms, thereby rendering the system non-integrable~\cite{Sachdev2011,Zamolodchikov1989}. Beyond the loss of integrability, the longitudinal field explicitly breaks the underlying $\mathbb{Z}_2$ symmetry, modifies the excitation spectrum, and stabilizes new magnetic phases that are absent in the integrable model. These effects lead to richer quantum critical behavior, enhanced many-body correlations, and confinement phenomena, making the longitudinal-field Ising model an important prototype for studying strongly interacting quantum systems beyond exactly solvable limits~\cite{Kormos2017,Sachdev2011}. Moreover, longitudinal fields naturally arise in several experimental platforms, including Rydberg atom arrays, trapped-ion simulators, and superconducting qubit architectures, where local control fields and device imperfections break the idealized symmetry of the transverse-field Ising model~\cite{Bernien2017,Kjaergaard2020,Monroe2021}.

\begin{figure*}[ht]
    \subfloat[\centering   ]{{\includegraphics[width=5cm]{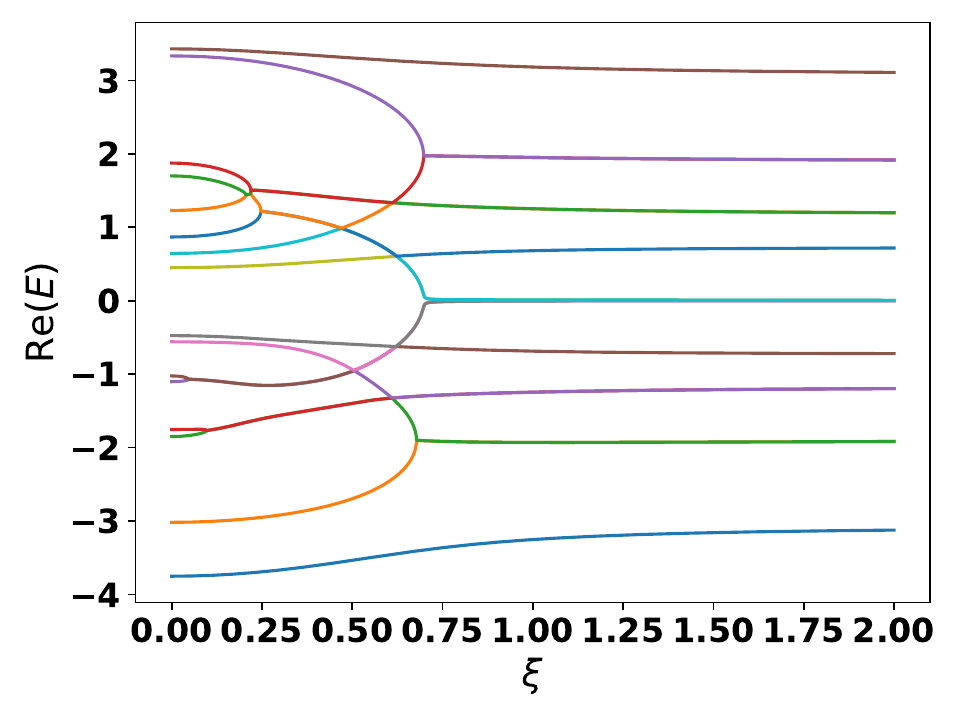} }}%
    \qquad
    \subfloat[\centering  ]{{\includegraphics[width=5cm]{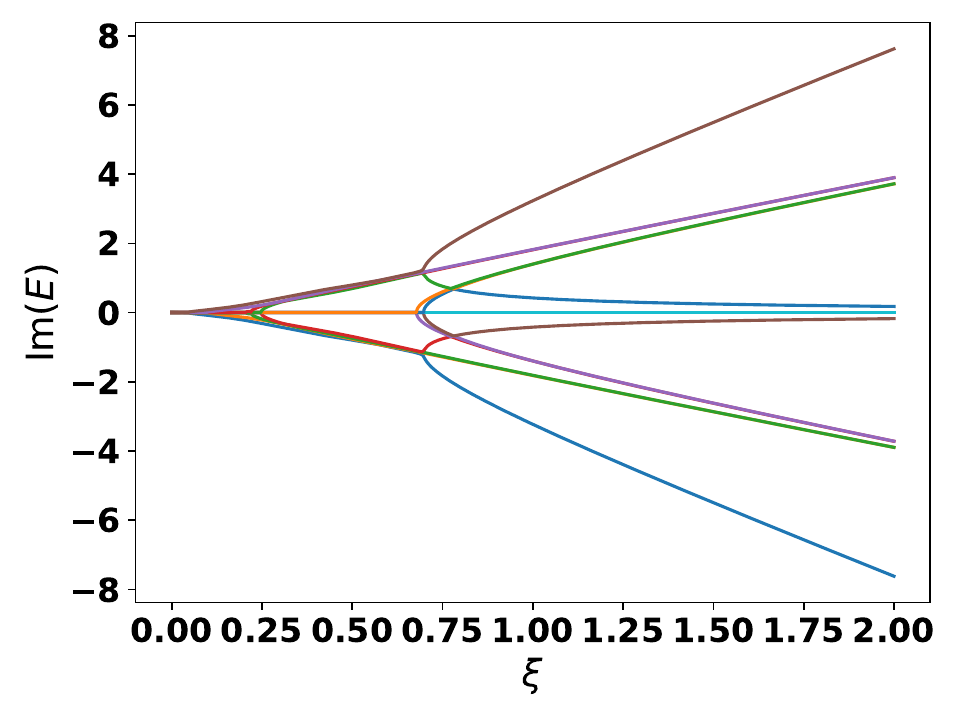} }}%
    \qquad
    \subfloat[\centering   ]{{\includegraphics[width=5cm]{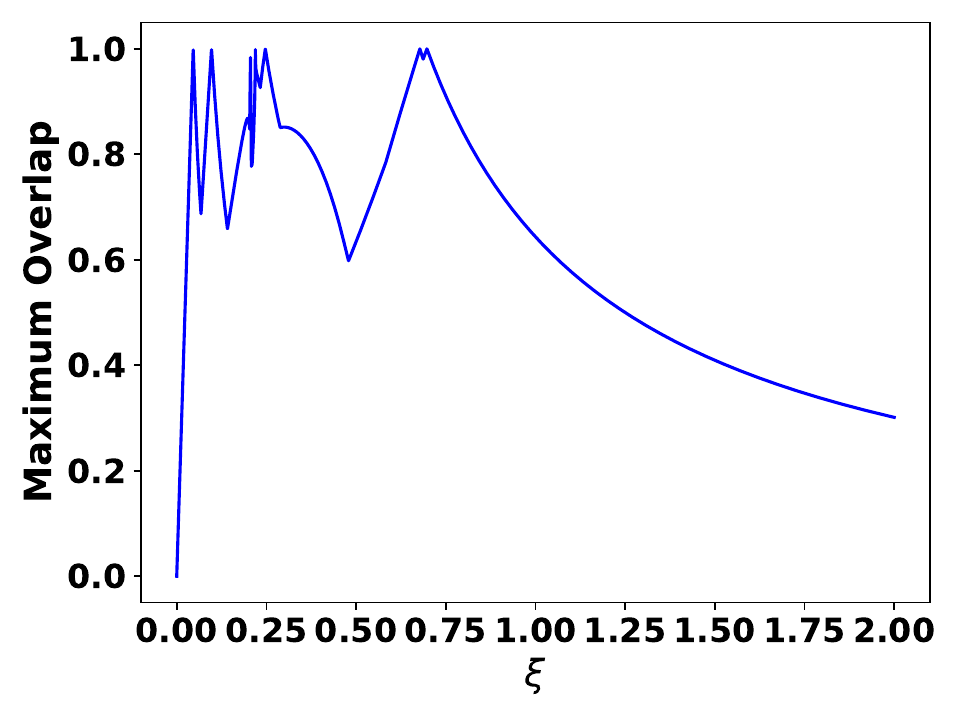} }}%
    \qquad
    \subfloat[\centering  ]{{\includegraphics[width=5cm]{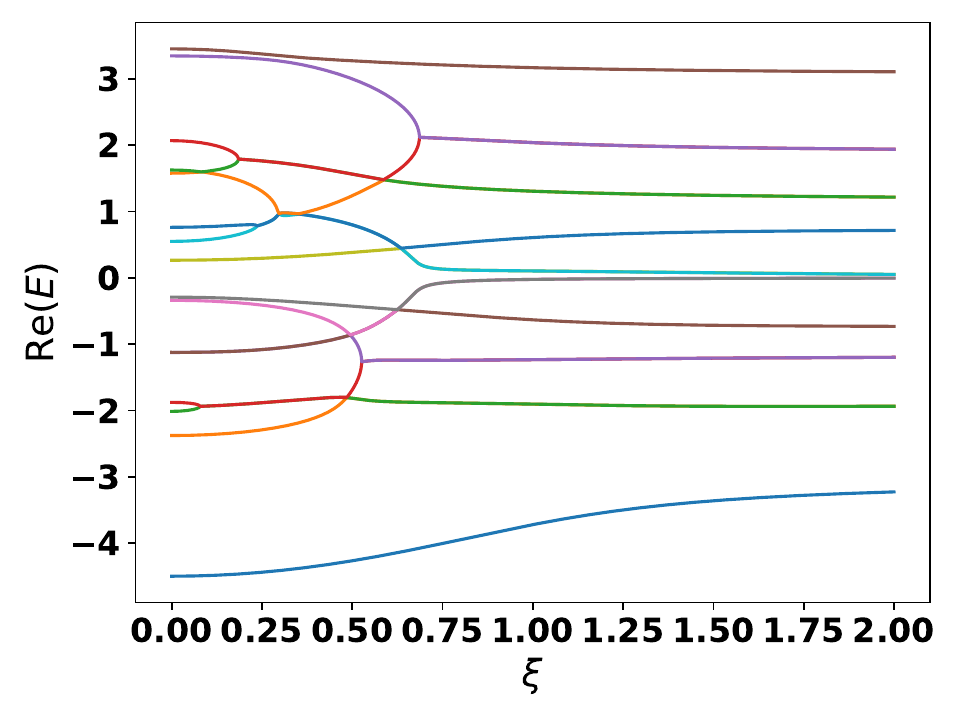} }}%
     \qquad
    \subfloat[\centering   ]{{\includegraphics[width=5cm]{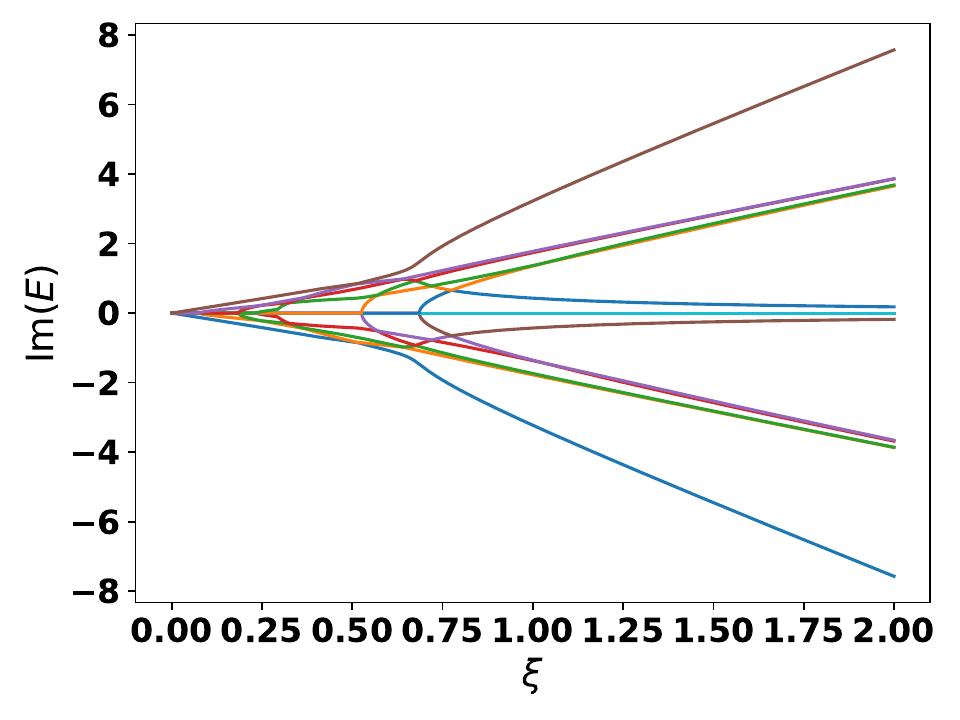} }}%
    \qquad
    \subfloat[\centering  ]{{\includegraphics[width=5cm]{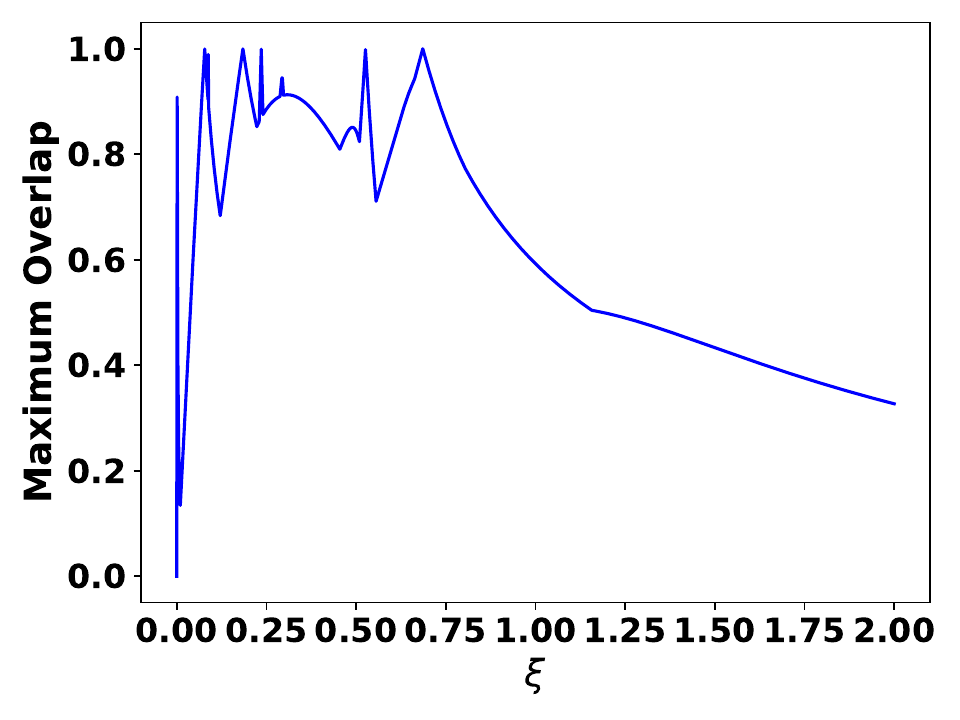} }}%
    \qquad
    \subfloat[\centering   ]{{\includegraphics[width=5cm]{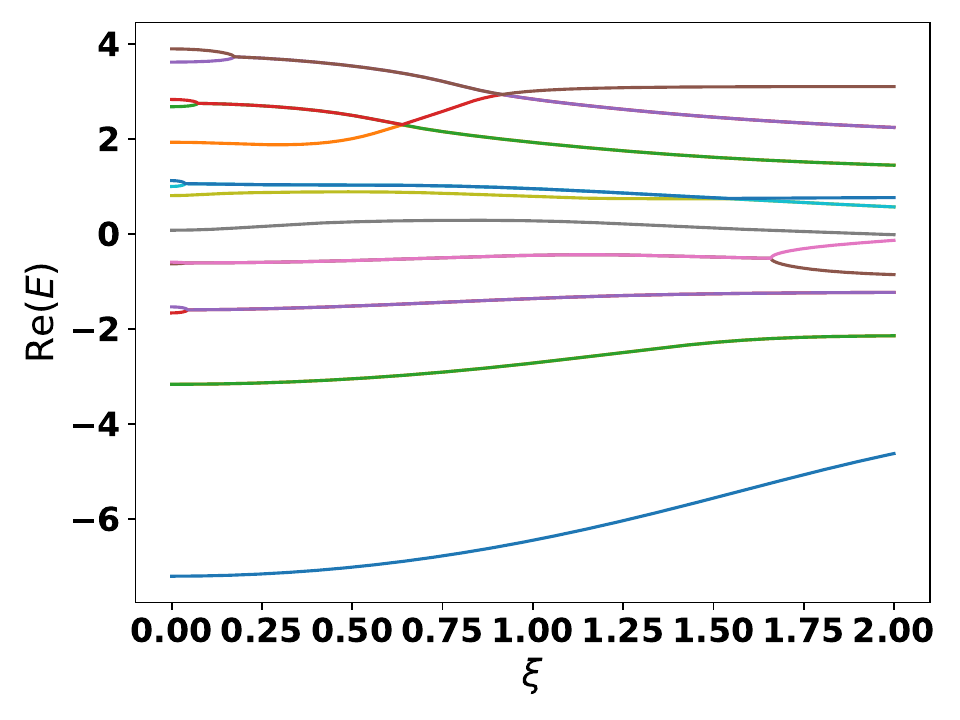} }}%
    \qquad
    \subfloat[\centering  ]{{\includegraphics[width=5cm]{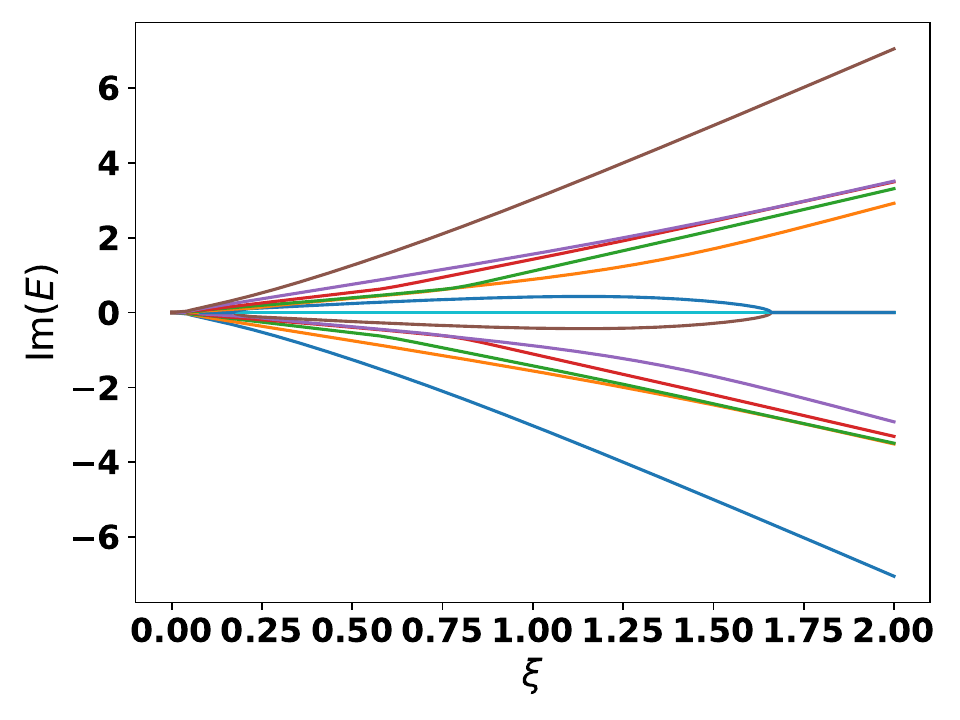} }}%
     \qquad
    \subfloat[\centering  ]{{\includegraphics[width=5cm]{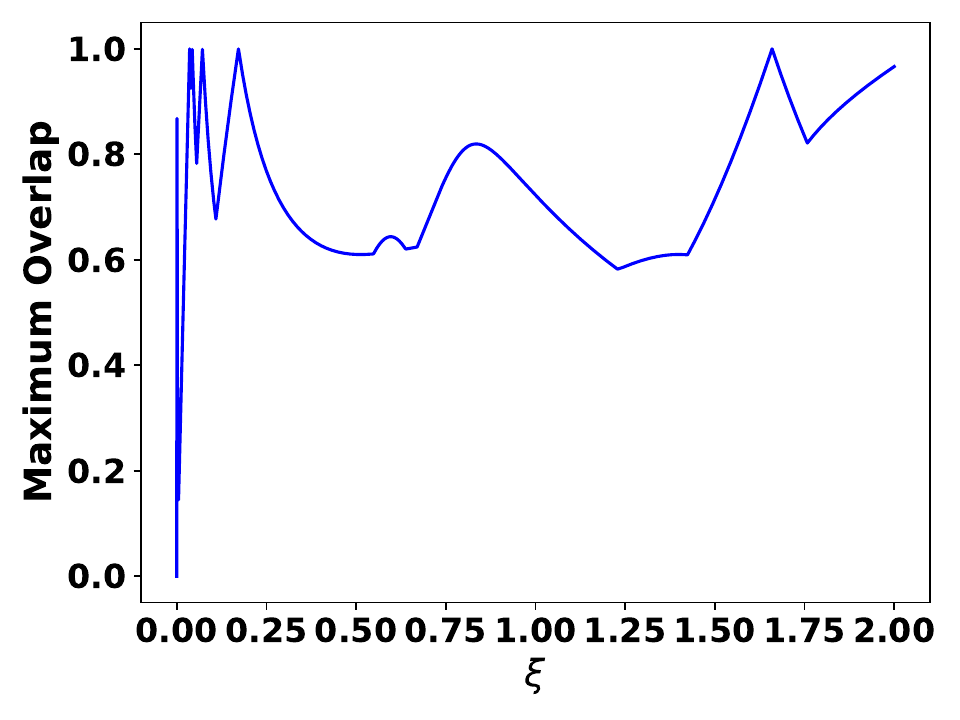} }}%
    
   \caption{\textbf{Symmetry breaking and exceptional points obtained with the ED.} Energy spectrum of the system for varying longitudinal fields $h$, illustrating the emergence of exceptional points (EPs). Columns display the real parts (\textbf{a}, \textbf{d}, \textbf{g}), imaginary parts (\textbf{b}, \textbf{e}, \textbf{h}), and the maximum overlap (\textbf{c}, \textbf{f}, \textbf{i}), respectively. Parameters are fixed at $N = 4$ and $\eta = 0.5$, with rows tracking field strengths of $h = 0.1$ (\textbf{a}--\textbf{c}), $h = 0.3$ (\textbf{d}--\textbf{f}), and $h = 1.0$ (\textbf{g}--\textbf{i}).}%
    \label{f1}%
\end{figure*}

In recent years, considerable attention has shifted toward non-Hermitian extensions of quantum many-body systems \cite{Ashida2020,Ding2022,Matsumoto2020}. Such systems arise naturally in open quantum dynamics, engineered dissipation, optical lattices with gain and loss, measurement-induced evolution, and superconducting quantum circuits. Unlike conventional Hermitian Hamiltonians, non-Hermitian systems possess complex-valued energy spectra and can exhibit exceptional points, non-reciprocal dynamics, and unconventional critical behavior\cite{Bender1998,Bender2007,Ashida2020,wah2026bridging,Bergholtz2021,lavoisier2025, wah2026autoregressive}.
Non-Hermiticity fundamentally modifies the spectral and dynamical properties of interacting quantum systems. In spin chains, complex external fields can induce effective gain and loss processes, alter quantum phase transitions, and generate new classes of non-equilibrium behavior. The interplay between coherent spin interactions, transverse quantum fluctuations, and non-Hermitian longitudinal or transverse fields therefore provides a rich framework for exploring novel quantum phenomena beyond standard equilibrium physics \cite{Lee2014,Kawabata2019,wah2026mbme, Hamazaki2019}.

Despite important analytical progress in special limits, the theoretical investigation of non-Hermitian interacting many-body systems remains highly challenging because the Hilbert-space dimension grows exponentially with system size, rendering exact diagonalization applicable only to relatively small systems. Tensor-network approaches, particularly matrix product states (MPS), have achieved remarkable success in one-dimensional quantum many-body systems by providing efficient representations of states with limited entanglement \cite{Schollwock2011,Orus2019,Cirac2021}. However, their efficiency is closely tied to the entanglement structure of the target state, which may become restrictive in strongly correlated or highly entangled regimes. These limitations motivate the development of alternative variational approaches, such as neural-network quantum states (NQS), which provide flexible and scalable representations of many-body wave functions without imposing a fixed tensor-network geometry.

\begin{figure*}[ht]
    \subfloat[\centering   ]{{\includegraphics[width=5cm]{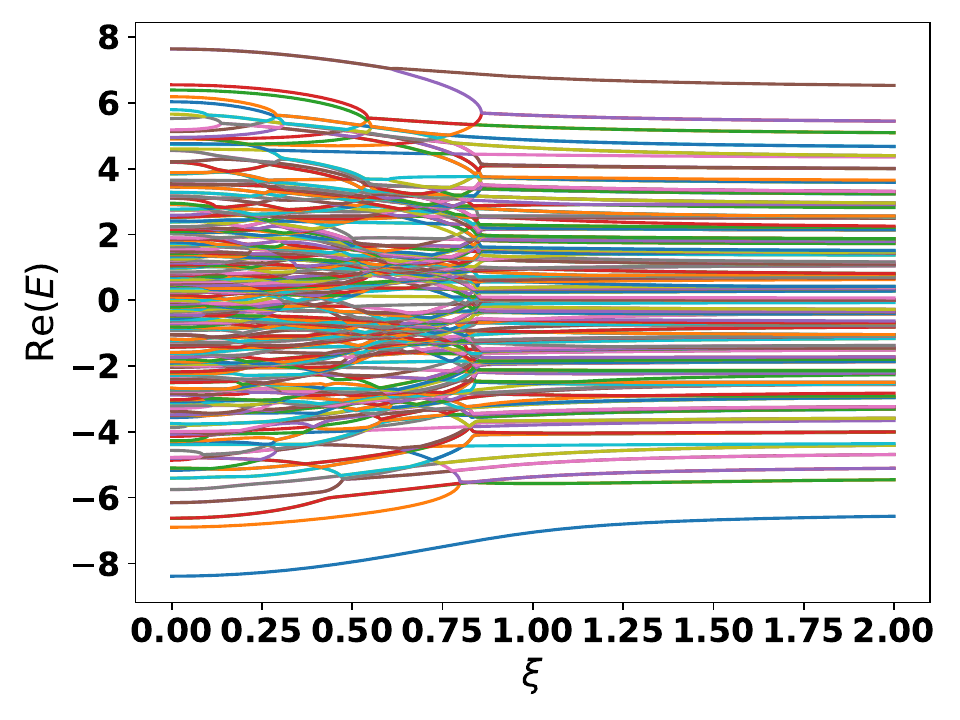} }}%
    \qquad
    \subfloat[\centering  ]{{\includegraphics[width=5cm]{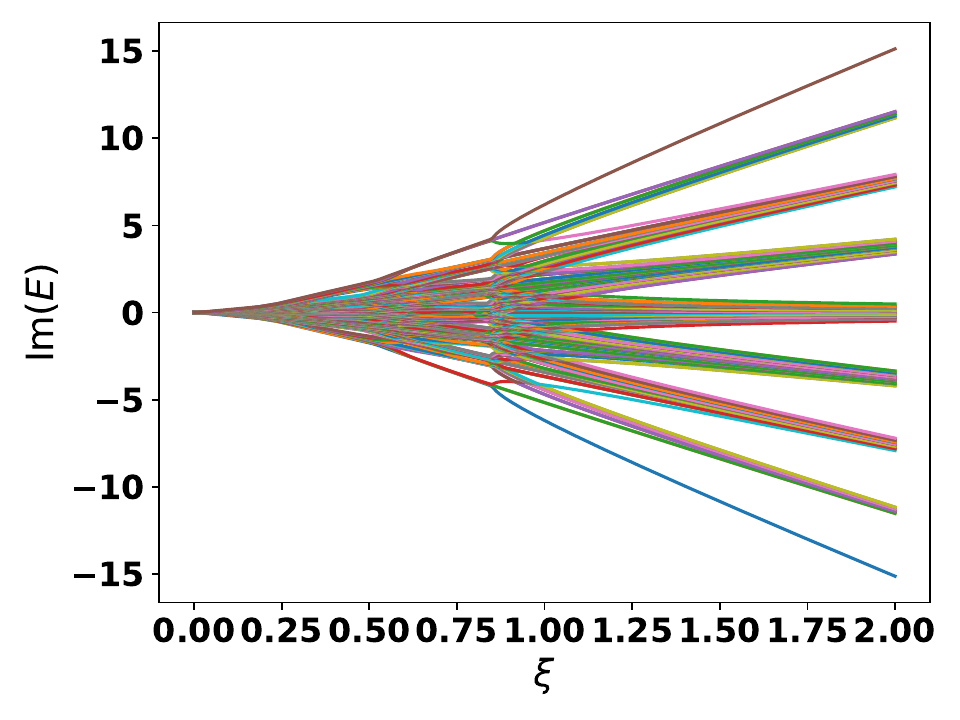} }}%
    \qquad
    \subfloat[\centering   ]{{\includegraphics[width=5cm]{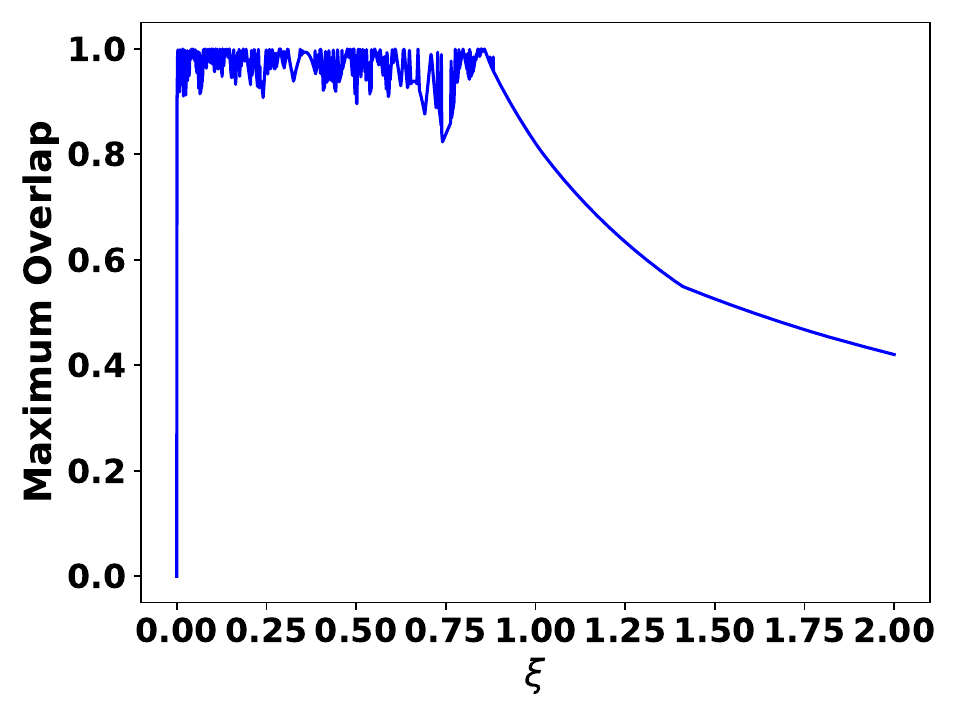} }}%
    \qquad
    \subfloat[\centering  ]{{\includegraphics[width=5cm]{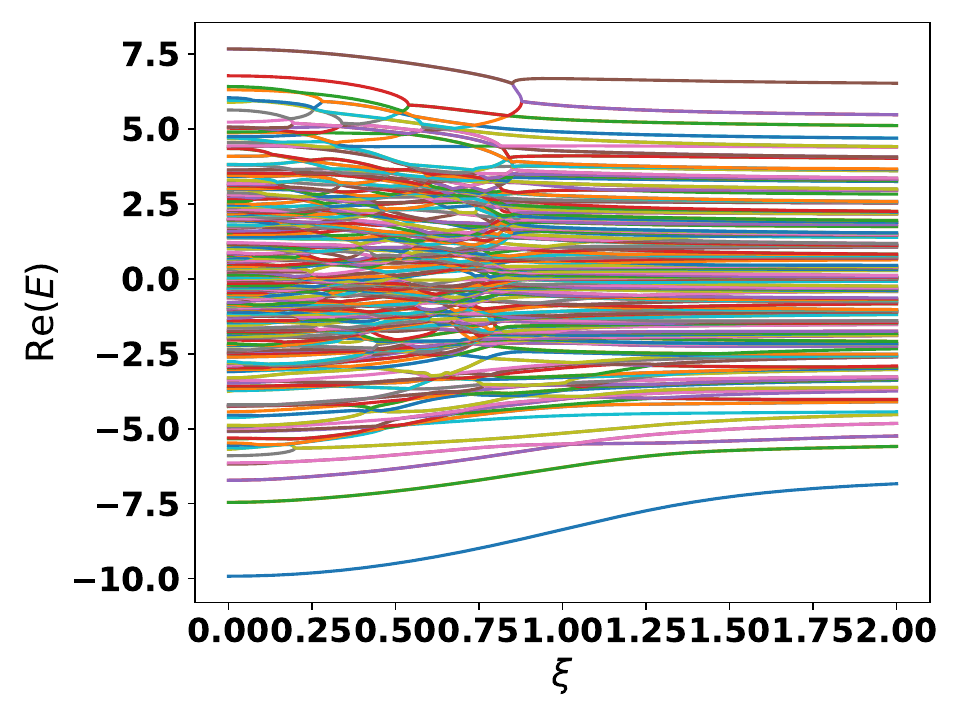} }}%
     \qquad
    \subfloat[\centering   ]{{\includegraphics[width=5cm]{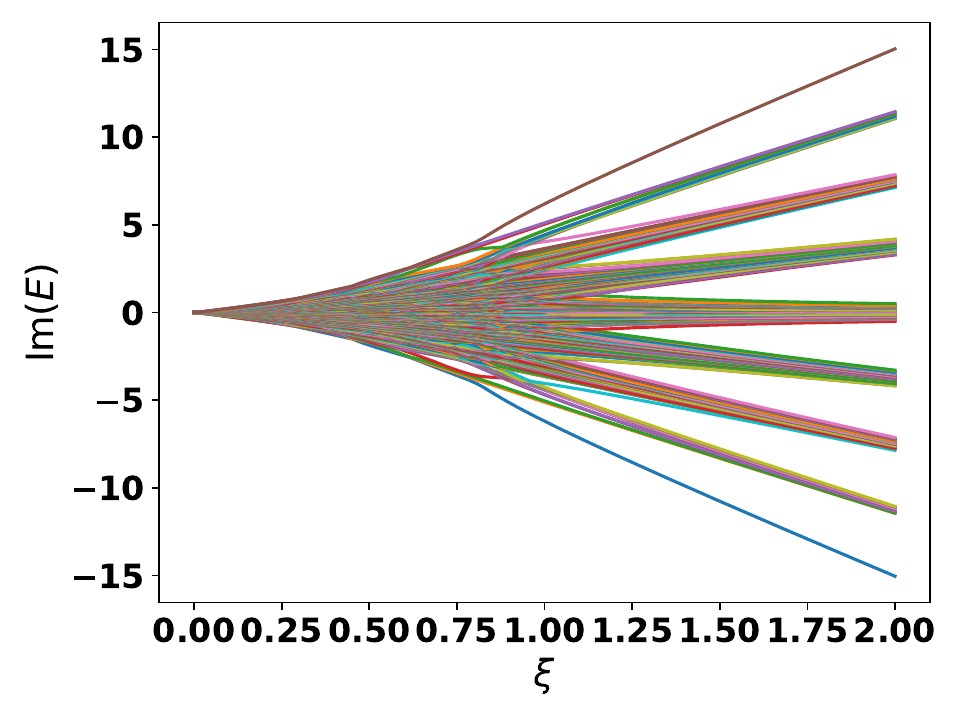} }}%
    \qquad
    \subfloat[\centering  ]{{\includegraphics[width=5cm]{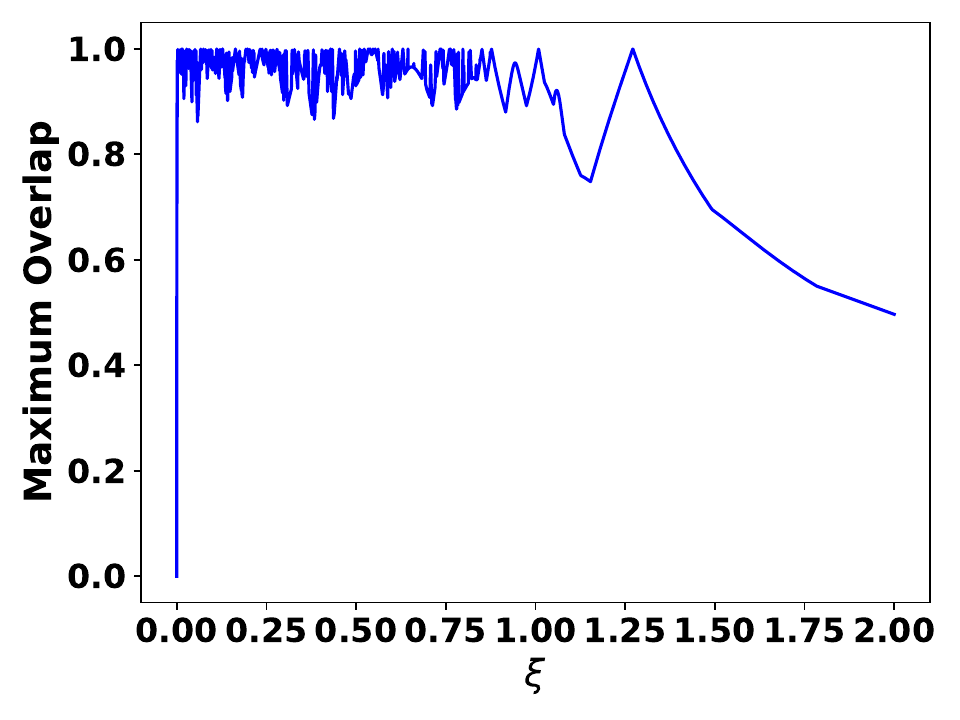} }}%
    \qquad
    \subfloat[\centering   ]{{\includegraphics[width=5cm]{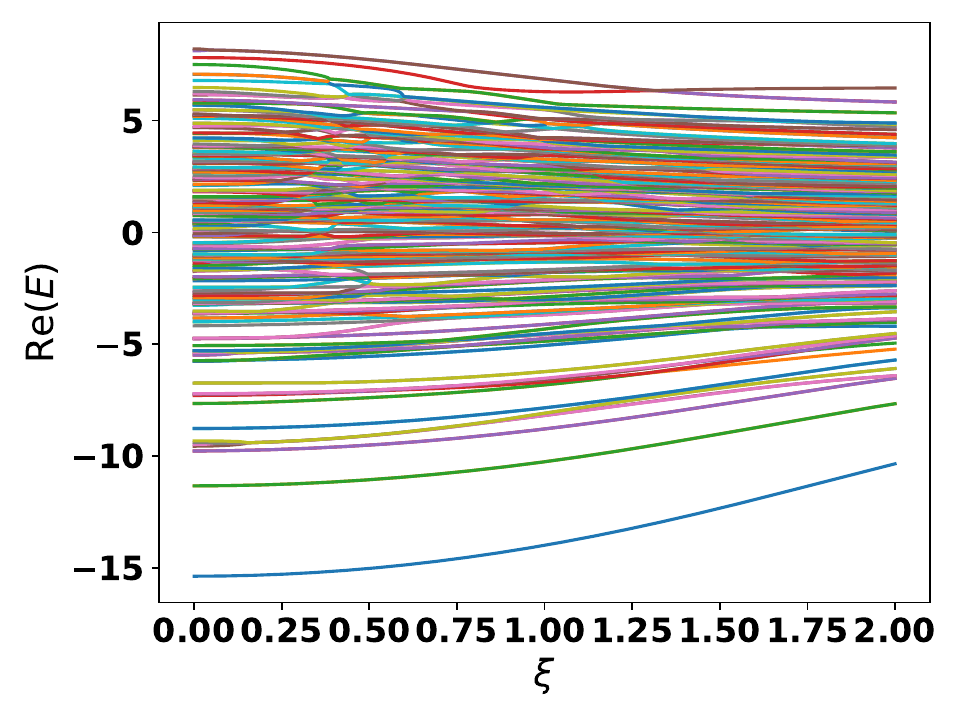} }}%
    \qquad
    \subfloat[\centering  ]{{\includegraphics[width=5cm]{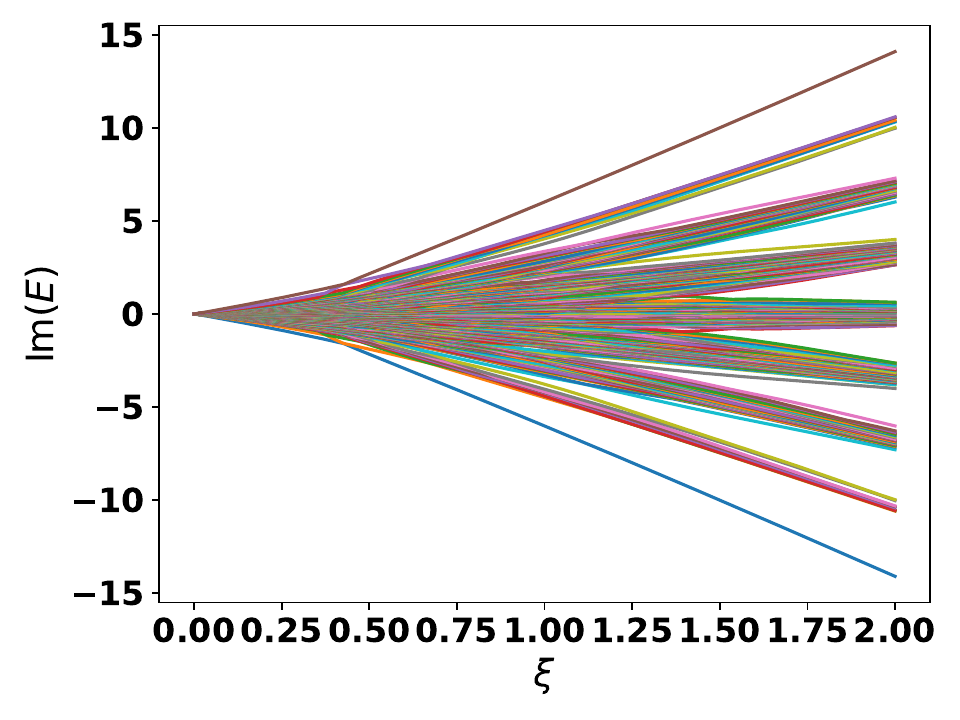} }}%
     \qquad
    \subfloat[\centering  ]{{\includegraphics[width=5cm]{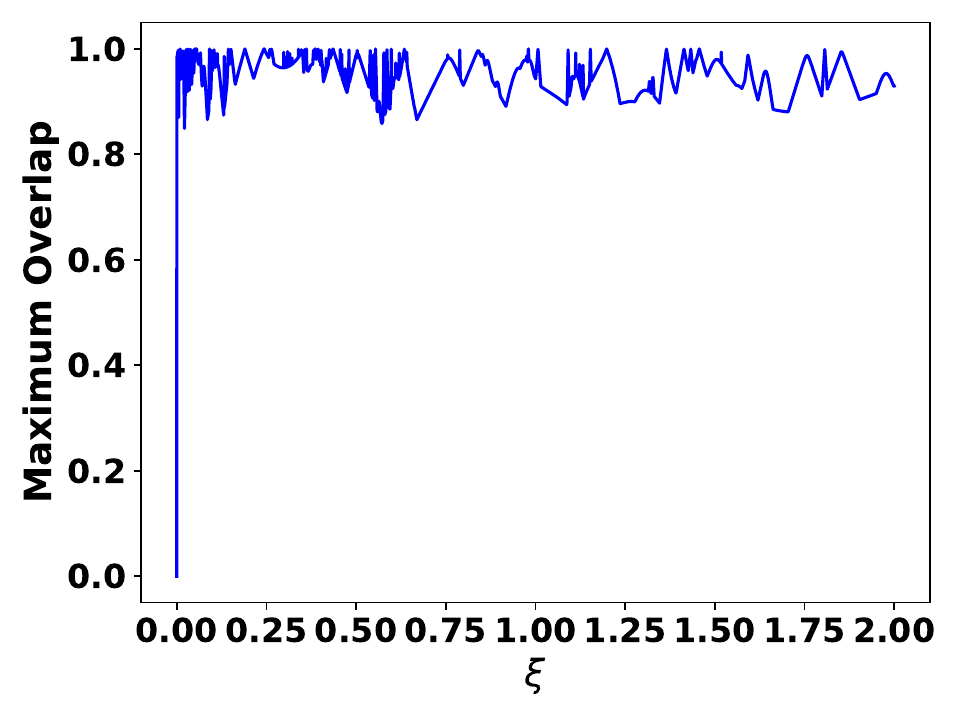} }}%
    
       \caption{\textbf{Symmetry breaking and exceptional points.} Energy spectrum of the system for varying longitudinal fields $h$, illustrating the emergence of exceptional points (EPs). Columns display the real parts (\textbf{a}, \textbf{d}, \textbf{g}), imaginary parts (\textbf{b}, \textbf{e}, \textbf{h}), and the maximum overlap (\textbf{c}, \textbf{f}, \textbf{i}), respectively. Parameters are fixed at $N = 8$ and $\eta = 0.5$, with rows tracking field strengths of $h = 0.1$ (\textbf{a}--\textbf{c}), $h = 0.3$ (\textbf{d}--\textbf{f}), and $h = 1.0$ (\textbf{g}--\textbf{i}). Band coloring corresponds to their chronological order of appearance.}%
    \label{f2}%
\end{figure*}

Neural quantum states (NQSs) based on RBMs provide a flexible  expressive variational framework for representing the quantum many-body wave function of NH systems as initially introduced in Ref.\cite{lavoisier2025}. Unlike conventional RBMs naturally accommodate complex-valued amplitudes and phases while efficiently encoding highly non-local many-body correlations through a relatively small number of variational parameters~\cite{Carleo2017,Deng2017,Melko2019,Carrasquilla2023,Glasser2018,Nomura2021}. Combined with Variational Monte Carlo (VMC), they offer a scalable computational approach that enables the study of quantum systems well beyond the size limitations of exact diagonalization.

In this work, we make four main contributions. First, we formulate a RBM neural quantum-state framework for the non-Hermitian transverse-field Ising chain in the presence of an explicit longitudinal magnetic field, thereby extending previous neural-network approaches to a non-integrable interacting many-body system. Second, we benchmark the RBM variational Monte Carlo (VMC) method against exact diagonalization for finite lattice sizes, demonstrating its high accuracy in reproducing the ground-state energy and magnetic observables. Third, we investigate how the interplay between transverse quantum fluctuations, longitudinal interactions, and non-Hermiticity modifies the ground-state properties, magnetization, spin-spin correlations, and the parity-time ($\mathcal{PT}$) symmetry-breaking transition. Finally, we demonstrate that the proposed neural-network approach scales efficiently to larger system sizes beyond the reach of exact diagonalization, providing a powerful computational framework for exploring quantum criticality and emergent magnetic order in interacting non-Hermitian quantum many-body systems~\cite{Ashida2020,Bergholtz2021}.

Although neural-network quantum states (NQSs) have recently been employed to investigate non-Hermitian Ising models~\cite{lavoisier2025, wah2026bridging, wah2026autoregressive}, existing studies have primarily focused on the standard non-Hermitian transverse-field Ising chain without an explicit longitudinal magnetic field. The introduction of a longitudinal field fundamentally changes the problem by destroying integrability, explicitly breaking the underlying $\mathbb{Z}_2$ symmetry, modifying the exceptional-point (EP) structure, and generating new magnetic ordering phenomena that cannot be captured within the integrable model. The present work therefore extends the neural-network quantum-state framework to a substantially more challenging interacting non-Hermitian many-body system and demonstrates that RBM wave functions accurately characterize the parity-time ($\mathcal{PT}$)-symmetry-breaking transition and the emergence of magnetic order.

The remainder of this work is organized as follows. Sec.\ref{part2}, we introduce the non-Hermitian Hamiltonian and describe its structural and parity-time ($\mathcal{PT}$) symmetry-breaking properties. In Sec.\ref{part3}, we detail the neural-network variational ansatz based on complex-valued RBMs and the optimization routine. Sec.\ref{part4}presents our numerical findings, benchmarking the RBM results against exact diagonalization and mapping out the quantum phase transition via energy scaling and longitudinal magnetization profiles. Finally, concluding remarks are provided in Sec.\ref{part5}.

\section{Model Hamiltonian and Description}\label{part2}

We consider a non-Hermitian extension of the one-dimensional
transverse-field Ising model incorporating both transverse and
longitudinal magnetic fields. The system is defined on a chain of
$N$ spin-$1/2$ particles with periodic boundary conditions
$\sigma_{N+1}=\sigma_1$ and is governed by the Hamiltonian \cite{lavoisier2025,Yang2017, wah2026bridging}:

\begin{equation}\label{E1}
H = -\sum_{j=1}^{N} J_z \sigma_j^z \sigma_{j+1}^z - \sum_{j \in A} g_j \sigma_j^x - \sum_{j \in B} g_j^* \sigma_j^x - h\sum_j \sigma_j^z
\end{equation}

Here, $\sigma_j^z$ and $\sigma_j^x$ are Pauli operators acting on site $j$. $J_z$ denotes the strength of the nearest-neighbor ferromagnetic interaction, $g_j$ is the transverse magnetic field, and $h$ is the longitudinal magnetic field. The overall amplitude of the transverse field is taken to be complex, $g_j = \eta + i \xi$,  where $\eta$ and $\xi$ are real parameters. 
The real part of $g_j$ controls the conventional coherent spin-flip dynamics, while the imaginary part introduces non-Hermiticity associated with effective gain/loss processes.

\subsection{Hermitian and non-integrable limits}

The Hamiltonian in Eq.~(\ref{E1}) continuously interpolates
between several physically important limits.

In the Hermitian limit, obtained by setting the imaginary part of
the transverse field to zero ($\xi=0$), the Hamiltonian
reduces to the conventional transverse field Ising model in the
presence of a longitudinal magnetic field. When the longitudinal
field is further removed ($h=0$), the model becomes the exactly
solvable transverse-field Ising chain, which can be mapped onto
non-interacting fermions through the Jordan--Wigner
transformation followed by a Bogoliubov transformation
\cite{Lieb1961,Pfeuty1970}.

The introduction of a finite longitudinal magnetic field
fundamentally changes the problem. It destroys the free-fermion
structure by generating interacting fermionic terms, thereby
rendering the model non-integrable. Simultaneously, the
longitudinal field explicitly breaks the global
$\mathbb{Z}_2$ spin-flip symmetry of the conventional
transverse-field Ising model, lifting the degeneracy between the
two ferromagnetic ground states and modifying the nature of the
quantum phase transition
\cite{Sachdev2011,Zamolodchikov1989,Kormos2017}.

The simultaneous presence of transverse and longitudinal magnetic fields enriches the physical behavior of the system. The transverse field tends to align the spins along the $x$ direction and induces quantum fluctuations, whereas the longitudinal field favors alignment along the interaction axis ($z$ axis) and explicitly breaks the $\mathbb{Z}_2$ symmetry of the conventional transverse-field Ising model. Their competition gives rise to nontrivial many-body dynamics and significantly modifies the quantum critical behavior of the chain \cite{Pfeuty1970,Sachdev2011,Dutta2015}.

\subsection{PT symmetry and exceptional points}

 A non-Hermitian Hamiltonian can possess an entirely real eigenvalue spectrum provided it commutes with the combined parity-time operator, $[\mathcal{PT},H]=0$, these Hamiltonians are said to be $\mathcal{PT}$-symmetric \cite{Bender1998,Bender2007,ATEUAFACK2025130216}. Here, the parity operator $\mathcal{P}$ performs spatial reflection of the lattice sites, $\mathcal{P}j\mathcal{P}^{-1}=N-j+1$, while the time-reversal operator $\mathcal{T}$ performs complex conjugation, $\mathcal{T} i \mathcal{T}^{-1}=-i$. Under the combined transformation $\mathcal{PT}$, the imaginary contribution to the transverse field changes sign through complex conjugation, while spatial reflection preserves the lattice structure.

In the regime where $\mathcal{PT}$ symmetry remains unbroken, the energy spectrum remain entirely real despite the non-Hermitian nature of the Hamiltonian. However, beyond a critical non-Hermitian strength, spontaneous $\mathcal{PT}$-symmetry breaking occurs, leading to complex-conjugate eigenvalue pairs and qualitative modifications of the system dynamics and critical properties \cite{Bender1998,Bender2007,ElGanainy2018,Ashida2020,ATEUAFACK2025130216}.

The study of $\mathcal{PT}$-symmetric non-Hermitian spin chains is therefore of particular interest because it provides a controlled framework for exploring the transition between Hermitian and genuinely non-Hermitian quantum phases. In interacting many-body systems, this transition may strongly influence magnetization behavior, correlation functions, entanglement properties, and dynamical responses \cite{Ashida2020,Bergholtz2021,Lee2014,Yang2017}.

To analyze the role of system size, Fig.~\ref{f2} presents the energy spectrum for $N = 8$, revealing a structural proliferation of EPs at four different values of $\xi$.

 Fig.(\ref{f1})  illustrates the energy spectrum for a small chain ($N=4$) under varying longitudinal fields $h$, highlighting the emergence of exceptional points (EPs)—singularities where eigenvalues and their corresponding eigenvectors simultaneously coalesce. As shown in Fig.(\ref{f2}), increasing the system size to $N=8$ leads to structural proliferation of EPs in momentum space. Crucially, the introduction of a non-zero longitudinal field $h$ deforms the complex Riemann surfaces of the energy spectrum, generating a negative effect on the stability of the EPs by shifting their onset and structurally altering the boundaries of the unbroken $\mathcal{PT}$-symmetric phase.

\begin{figure}
    \centering
    \includegraphics[width=9cm]{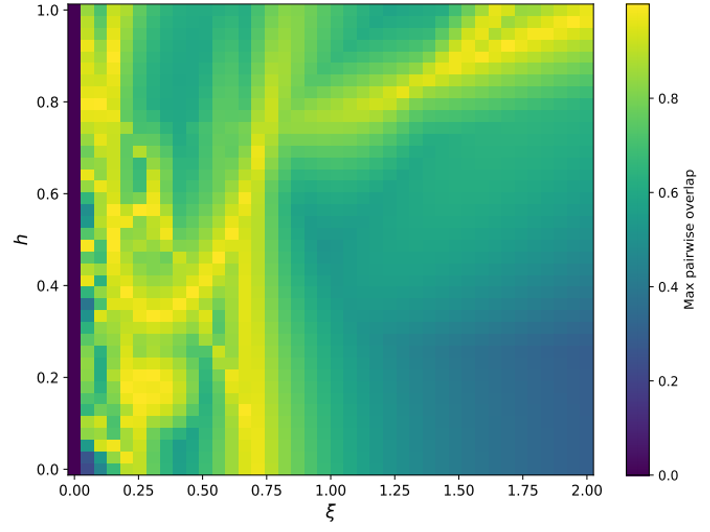} 
    \caption{\textbf{Maximum-overlap phase diagram in the $(h)-(\xi)$ parameter space.} The yellow regions identify the parameter regimes where exceptional points (EPs) occur, characterized by the coalescence of eigenstates and a maximum overlap equal to unity. The remaining parameters are fixed at $N=4$ and $\eta=0.5$.}
\label{f2a}%
\end{figure}

Fig.(\ref{f2a})  illustrates the interplay between the longitudinal field ($h$) and the imaginary part ($\xi$) of the transverse field, providing a map of the exceptional-point (EP) structure of the non-Hermitian system. The yellow patterns correspond to parameter values for which the maximum overlap reaches unity, signaling the coalescence of eigen states and hence the occurrence of an EP. Two qualitatively distinct regimes emerge: (a) a low regime ($\xi$), ($\xi\lesssim 1$) and (b) a high regime ($\xi$) ($\xi \gtrsim 1$).

In the low-($\xi$) regime, the EP region progressively shrinks as the longitudinal field (h) increases, indicating that the longitudinal field tends to suppress the parameter range over which the eigenstates coalesce. In contrast, in the high ($\xi$) regime, increasing ($h$) enhances the EP region, revealing a opposite response of the non-Hermitian spectrum to the longitudinal field. This qualitative reversal demonstrates that the effect of ($h$) on the symmetry structure is strongly controlled by the strength of the imaginary transverse field ($\xi$).

The resulting EP map therefore provides a direct characterization of the competition between the longitudinal field and non-Hermiticity. In particular, the change in the EP topology around ($\xi\sim1$) signals a qualitative reorganization of the spectral structure and the associated symmetry-breaking behavior. Thus, tuning ($h$) and ($\xi$) provides a means of controlling the location and extent of the EP regions, allowing the different symmetry regimes and their associated transitions to be systematically tracked.

\section{Neural-Network variational Ansatz}\label{part3}

To investigate these phenomena beyond the reach of exact diagonalization, we employ Restricted Boltzmann Machines (RBMs) optimized within a Variational Monte Carlo (VMC) framework. Within this neural-network quantum-state (NQS) formalism, the many-body wave function is parameterized using a discrete network architecture, as originally introduced by Carleo and Troyer~\cite{Carleo2017}. In contrast to approaches based on complex-valued neural-network parameters, we adopt a real-valued RBM ansatz throughout the present work, provided that the system remains $\mathcal{PT}$-symmetric. This choice provides a computationally simple and stable variational representation while allowing us to investigate the ground-state properties and magnetic observables of the non-Hermitian Hamiltonian considered here.

In general, because the underlying Hamiltonian $\hat{H}$ is non-Hermitian ($\hat{H}\neq\hat{H}^{\dagger}$), its left and right eigenstates, denoted by $\langle\Psi_L|$ and $|\Psi_R\rangle$, are no longer related by Hermitian conjugation~\cite{wah2026bridging,lavoisier2025, wah2026autoregressive}. Consequently, a complete variational description of non-Hermitian systems requires the independent optimization of two network configurations, $\mathbf{\Theta}_L$ and $\mathbf{\Theta}_R$, through Variational Monte Carlo sampling. The corresponding generalized variational energy is defined as
\begin{equation}\label{E2}
    E_{\text{var}}(\mathbf{\Theta}_L, \mathbf{\Theta}_R) = \frac{\langle \Psi_L(\mathbf{\Theta}_L) | \hat{H} | \Psi_R(\mathbf{\Theta}_R) \rangle}{\langle \Psi_L(\mathbf{\Theta}_L) | \Psi_R(\mathbf{\Theta}_R) \rangle},
\end{equation}
where  $\langle \Psi_L(\mathbf{\Theta}_L)|$ and $|\Psi_R(\mathbf{\Theta}_R)\rangle$  denote the parameterized left and right trial wave functions of $\hat{H}$. Owing to the biorthogonal nature of non-Hermitian systems, these states satisfy the biorthonormality condition
\begin{equation}\label{E3}
    \langle \Psi_{L, m}(\mathbf{\Theta}_L) | \Psi_{R, n}(\mathbf{\Theta}_R) \rangle = \delta_{mn}.
\end{equation}

\begin{figure*}[ht]
    \subfloat[\centering   ]{{\includegraphics[width=7.5cm]{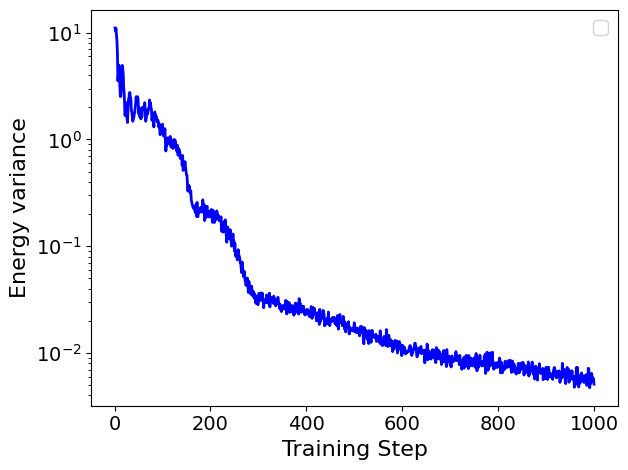} }}%
    \qquad
    \subfloat[\centering  ]{{\includegraphics[width=9.5cm]{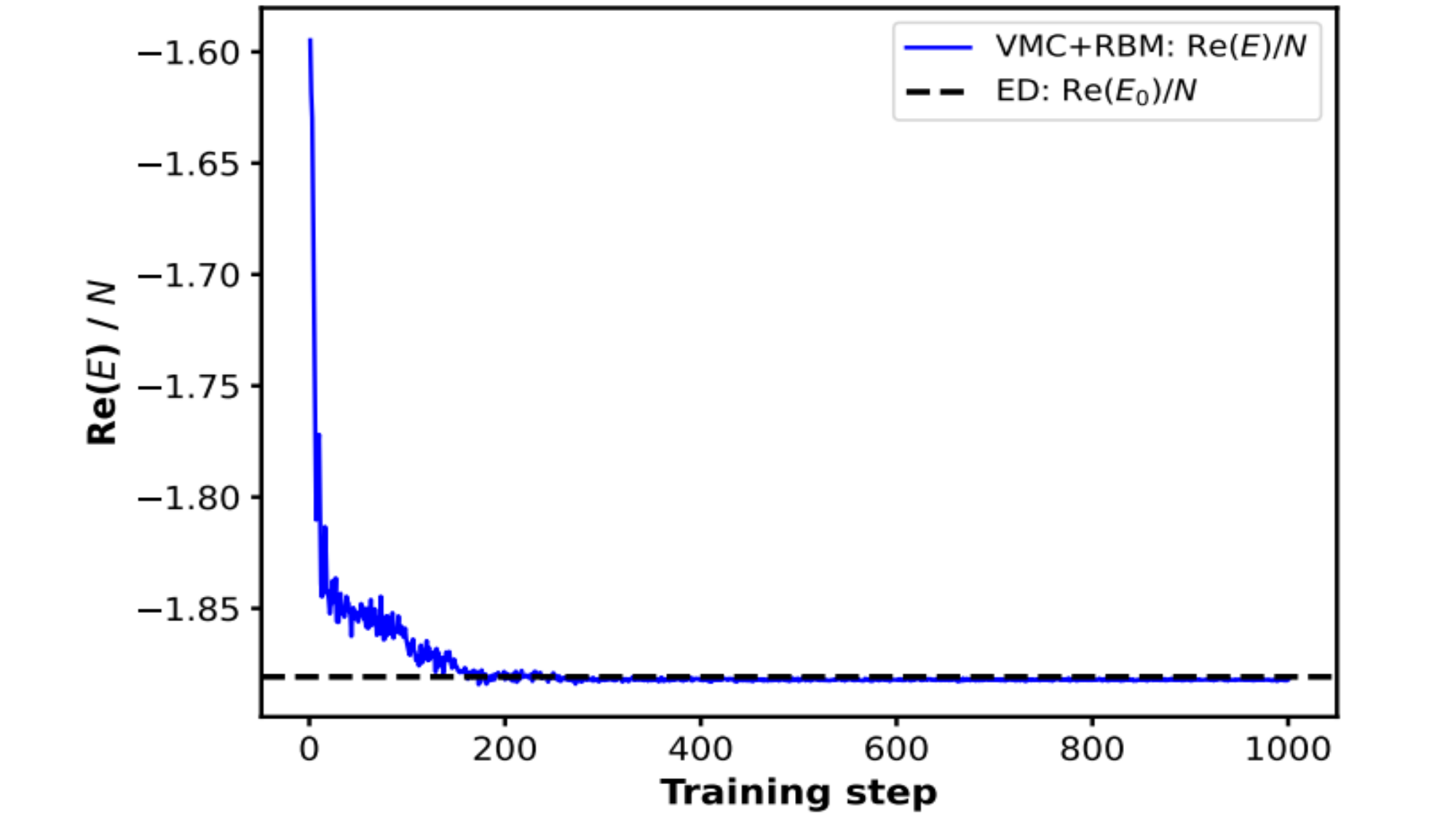} }}%
   \caption{\textbf{RBM vs. exact diagonalisation.}  (\textbf{a}) Energy variance of the real part of the local energy, and (\textbf{b}) the corresponding ground-state energy of the system as a function of training steps. The Restricted Boltzmann Machine (RBM) successfully converges to the exact ground-state energy predicted by exact diagonalization (indicated by the black dashed line). The variance curve demonstrates that convergence initiates around $200$ training steps. The fixed parameters for this simulation are $N = 10$ , $\eta = 0.5$, $\xi = 0.05$ and $h=0.1$.}%
    \label{f3}%
\end{figure*}

The above biorthogonal formulation provides the most general variational framework for non-Hermitian many-body systems. Nevertheless, Ref.~\cite{lavoisier2025} demonstrated that, for $\mathcal{PT}$-symmetric Hamiltonians in the unbroken $\mathcal{PT}$ phase, the ground-state optimization can be reformulated using a single real-valued neural-network ansatz without explicitly introducing independent left and right variational states. This simplification follows from the $\mathcal{PT}$ symmetry of the Hamiltonian, which preserves a real-valued representation of the variational ground state while retaining the correct physical observables. Consequently, the computational cost is significantly reduced by eliminating the simultaneous optimization of two independent neural networks. Motivated by this result, we adopt the real-valued RBM-VMC framework throughout the present work to investigate the ground-state properties and quantum critical behavior of the $\mathcal{PT}$-symmetric non-Hermitian Ising chain.

\subsection{Ground-State Properties from Restricted Boltzmann Machines}

In non-Hermitian quantum systems, the notion of a ground state is more subtle than in conventional Hermitian systems because the energy spectrum may be complex and the left and right eigenvectors are generally distinct \cite{Ashida2020,wah2026bridging}. Throughout this work, we define the ground state as the eigenstate whose eigenvalue possesses the smallest real part and which continuously connects to the Hermitian ground state as the non-Hermitian parameter is adiabatically introduced. This definition provides a physically meaningful characterization of the low-energy sector in both the $\mathcal{PT}$-symmetric and weakly broken $\mathcal{PT}$ phases \cite{wah2026bridging,Bender2007,lavoisier2025, wah2026autoregressive}. Consequently, our variational optimization is performed by minimizing the real part of the energy. However, we emphasize that the same framework can be readily adapted to target other spectral branches, including states with the largest real part or the largest/smallest imaginary component of the spectrum.

To represent the ground-state wave function $|\psi_{\theta,R}^{0}\rangle$, we employ a Restricted Boltzmann Machine (RBM) variational ansatz. The wave function is expressed in the computational basis, where a physical configuration is denoted by
$x=(x_1,x_2,\ldots,x_N)$ with $x_i\in\{-1,+1\}$. The RBM architecture consists of $N$ visible neurons representing the physical spins and $M$ auxiliary hidden neurons $\tilde{h}=(\tilde{h}_1,\tilde{h}_2,\ldots,\tilde{h}_M)$ with $\tilde{h}_j=\pm1$. The hidden layer is connected to the visible layer through complex variational parameters, while there are no intra-layer couplings \cite{Carleo2017,Deng2017,Nomura2021}.

After analytically summing the hidden variables, the amplitude of the RBM wave-function takes the form \cite{Melko2019}

\begin{equation}\label{E4}
\psi_{\theta,i}^0(x) = \langle x \vert \psi_{\theta,i}^0 \rangle=
\exp\left(\sum_{i=1}^{N}a_ix_i\right)
\prod_{j=1}^{M}
2\cosh\left(
b_j+\sum_{i=1}^{N}W_{ij}x_i
\right),
\end{equation}

where $a_i$, $b_j$, and $W_{ij}$ correspond to the visible biases, hidden biases, and visible-hidden coupling parameters, respectively. Because the Hamiltonian is non-Hermitian and contains the complex field parameter $g$, all variational parameters are chosen to be real-valued. This complex parameterization allows the RBM ansatz to efficiently encode both the probability amplitudes and the nontrivial phase structure of the ground-state wave function.
The variational parameters are optimized by minimizing the energy expectation value $E_{\text{var}} = \frac{\langle \psi_{L,\theta} \vert H \vert \psi_{R,\theta} \rangle}{\langle \psi_{L,\theta} \vert \psi_{R,\theta} \rangle}$ using the Variational Monte Carlo (VMC) approach. Within this framework, the energy is evaluated as a statistical average of the local energy \cite{lavoisier2025,wah2026bridging, wah2026autoregressive},
\begin{equation}\label{E5}
E_{\mathrm{loc}} = \frac{\langle x \vert H \vert \psi_{R,\theta}^0 \rangle}{\langle x \vert \psi_{R,\theta}^0 \rangle} = \frac{\langle \psi_{L,\theta}^0 \vert H \vert x \rangle}{\langle \psi_{L,\theta}^0 \vert x \rangle}
\end{equation}

It is worth emphasizing that the evaluation of local energy does not depend on whether the Hamiltonian acts on the left or right eigenstate. Within the present formulation, both approaches yield the same energy expectation value, as demonstrated in Ref.~\cite{lavoisier2025}. This property is particularly advantageous for non-Hermitian  $\mathcal{PT}$-symmetric systems because it avoids the explicit construction of a biorthogonal inner product, which would otherwise require the simultaneous optimization of two independent neural-network representations of the left and right eigenstates. Moreover, neither $(|\Psi_R\rangle)$ nor $(\langle\Psi_L|)$ needs to be individually normalized. Since Eq.~(\ref{E5}) involves only ratios of wave-function amplitudes, the local energy remains invariant under any global rescaling of the variational states.

Once the VMC optimization has converged, the resulting RBM parameters provide an accurate approximation of the many-body ground state. Ground-state observables, including the energy, magnetization, correlation functions, and entanglement-related quantities, can then be computed through Monte Carlo sampling of the optimized wave function. To assess the accuracy of the neural-network ansatz, the RBM results are systematically benchmarked against exact diagonalization calculations for finite system sizes.

\begin{figure}[ht]
    \centering
    \includegraphics[width=9cm]{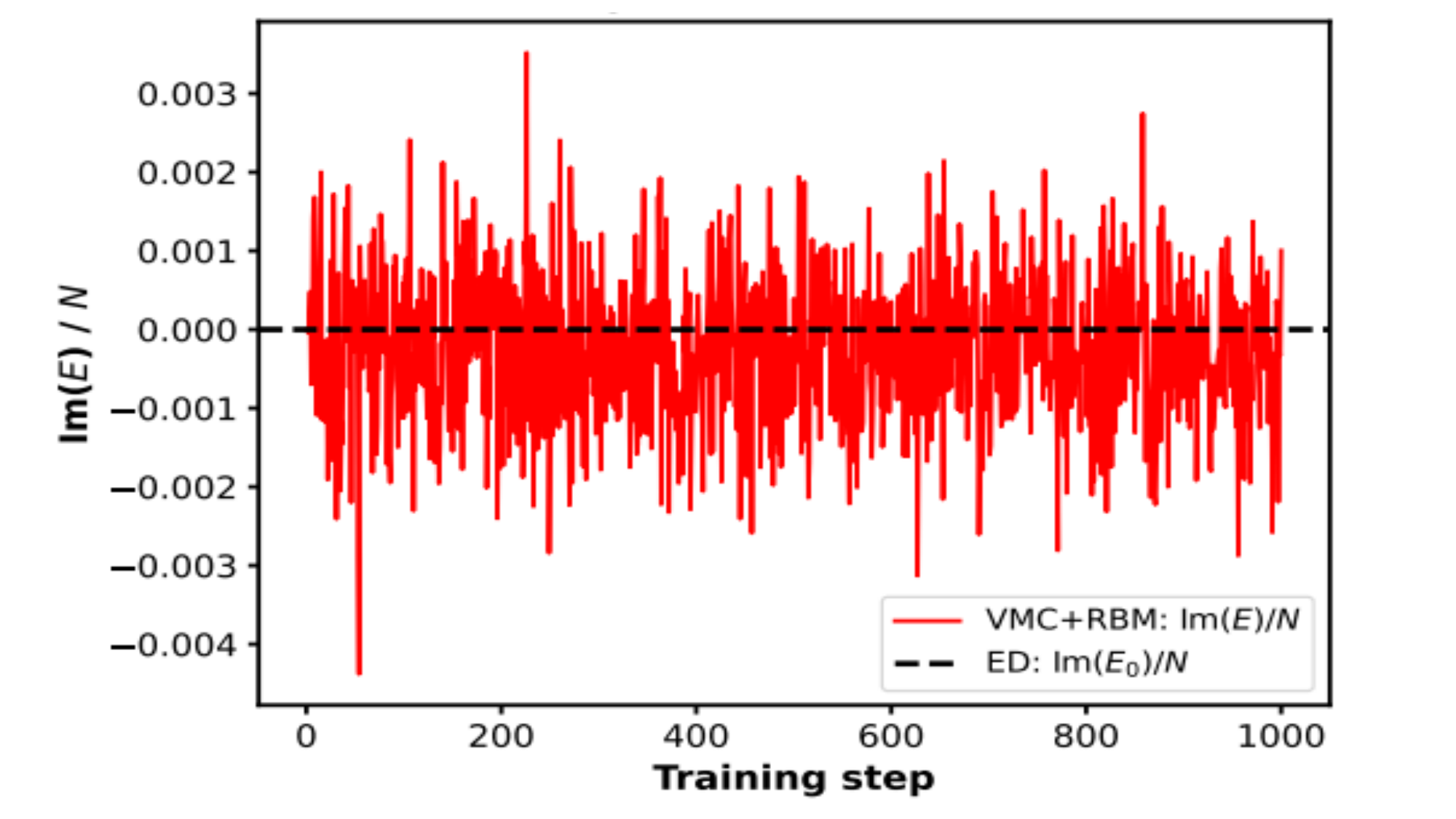} 
    \caption{\textbf{Imaginary part of the local energy}.The corresponding imaginary part of the local energy fluctuates around zero. energy of the system as a function of training steps. The fixed parameters for this simulation are $N = 10$ , $\eta = 0.5$, $\xi = 0.05$ and $h=0.1$. }%
    \label{f4}%
\end{figure}

We present the energy variance for the system in Fig.~\ref{f3}(a). The results indicate that the energy estimates exhibit minimal fluctuations and steadily converge toward the exact ground-state energy obtained via exact diagonalization (ED), as shown in Fig\ref{f3}(b). Furthermore, Fig.~\ref{f3}(b) tracks the evolution of the local energies throughout the Restricted Boltzmann Machine (RBM) training process. We observe that the RBM efficiently captures the ground-state properties of the system, a trend corroborated by the decreasing variance curve. 

In the spontaneously broken symmetry regime, the local energies computed during training generally take on complex values. However, the final 
ground-state energy estimate is retrieved through Monte Carlo averaging. Because these local energies predominantly appear in complex conjugate pairs, this averaging cancels out the imaginary components, yielding a purely real-valued result [cf. Fig.~\ref{f4}]. This mechanism is practically equivalent to extracting the real magnitude of the energy, as confirmed in Fig.~\ref{f3}(b), where the negative sign simply indicates that the energy of the ground-state possesses a negative real part. 

The close agreement between the real part of the averaged energy and its total magnitude suggests that the system operates in a regime where only a few states spontaneously break the $\mathcal{P}\mathcal{T}$ symmetry, positioning the system in close proximity to an exceptional point. Consequently, despite the broken symmetry at the local level, the global behavior retains a ``real-like'' spectrum, consistent with its proximity to the unbroken $\mathcal{P}\mathcal{T}$-symmetric phase in the regime of low longitudinal magnetic field $h$.

To capture the role of $h$ of the system we take the real part of the energy of the ground state.  

\begin{figure}[ht]
    \centering
    \includegraphics[width=8cm]{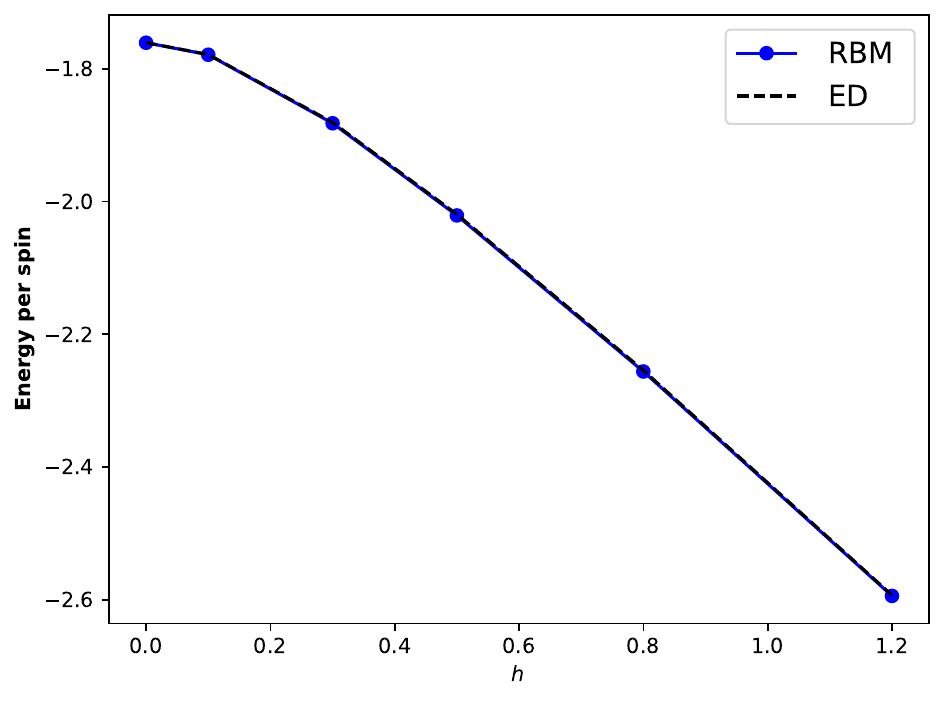} 
    \caption{\textbf{ground state energy vs. $h$}.The ground state energy provided by the RBM capture exactly  ED architecture for different values of the longitudinal field $h$.Parameter of the simulation are $N = 10$ , $\eta = 1.6$,and $\xi = 0.16$.}%
    \label{f5}%
\end{figure}
\begin{figure*}[ht]
    \subfloat[\centering   ]{{\includegraphics[width=8.5cm]{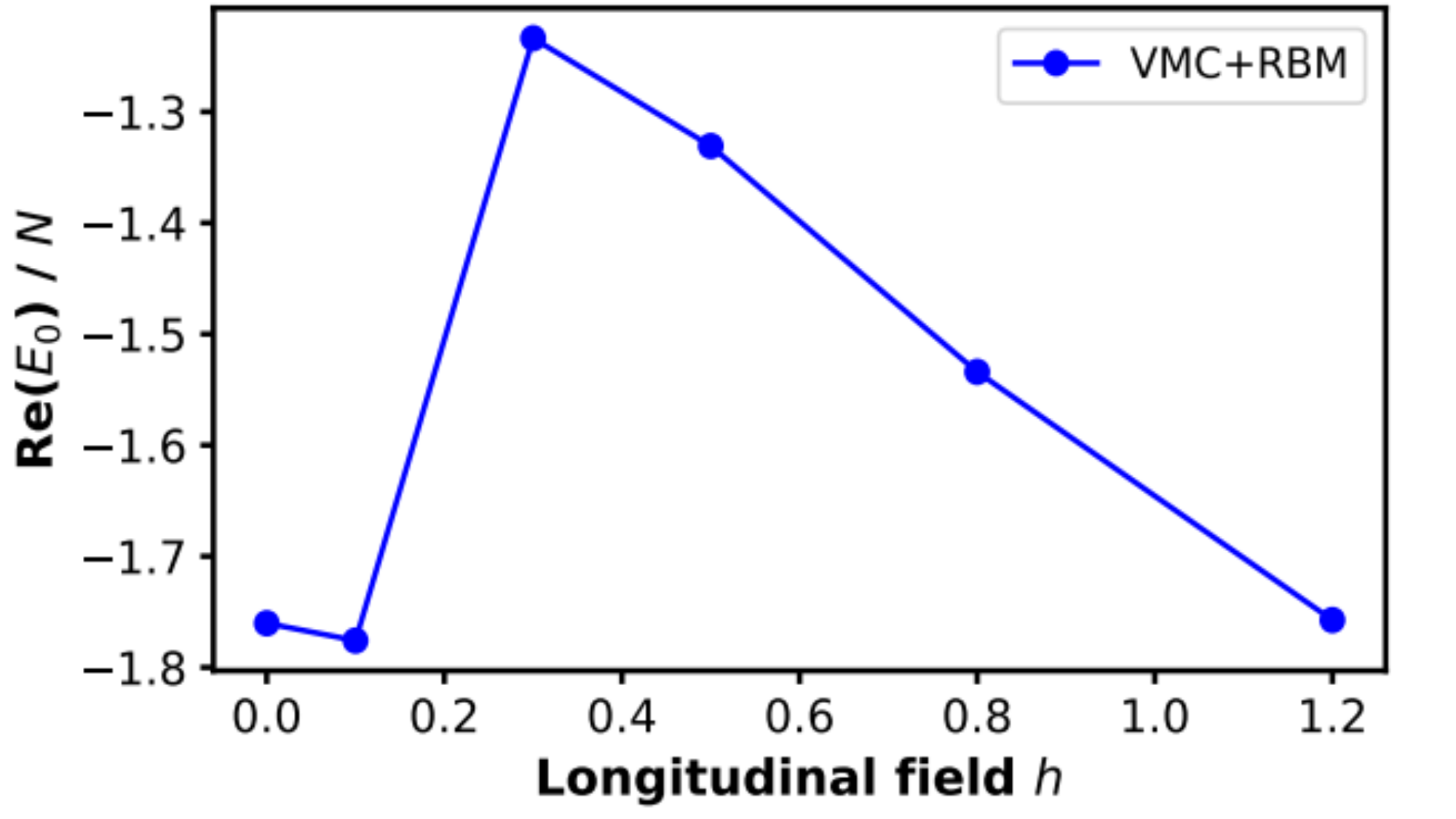} }}%
    \qquad
    \subfloat[\centering  ]{{\includegraphics[width=8.5cm]{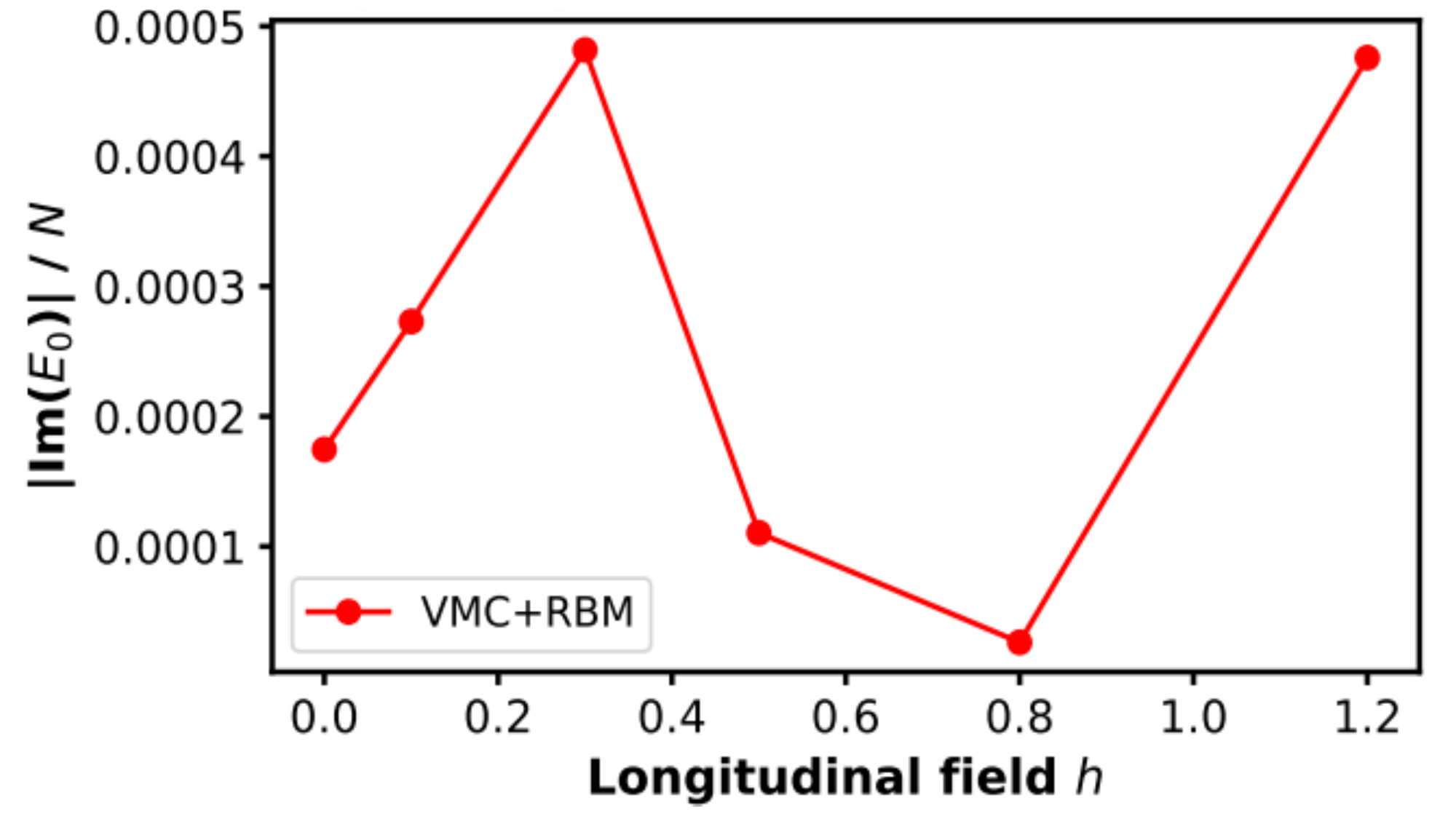} }}%
   \caption{\textbf{Ground state energy vs.$h$.}  (\textbf{a}) real part of the ground state energy, and (\textbf{b}) the corresponding imaginary part of ground-state energy of the system as a function of the longitudinal field $h$. The different $PT$ phase are well visible in (\textbf{b}).  The fixed parameters for this simulation are $N = 100$ , $\eta = 1.6$, $\xi = 0.16$. }%
    \label{f6}%
\end{figure*}

In addition, to capture the role played by the longitudinal field $h$, we investigate the energy of the ground-state per spin and construct the corresponding phase diagrams [cf. Fig.~\ref{f5} and Fig.\ref{f6}]. In the small-system-size regime ($N = 10$, Fig.~4), where ED is feasible, the RBM architecture performs exceptionally well, yielding results that closely match those obtained from ED. This confirms the ability of our protocol to effectively capture the 
system's quantum behavior. 

For larger system sizes beyond the reach of ED ($N = 100$, Fig.\ref{f6}), we present the corresponding phase diagram and observe that the system exhibits a spontaneous symmetry breaking around the critical value $h_c \approx 0.3$. This is clearly visible in Fig. \ref{f6}(a) via the real part of the ground-state energy. Despite the small magnitude of the imaginary component, this symmetry breaking strongly coexists in both the real and imaginary sectors at the same critical value, as shown in Fig.~\ref{f6}(b). This localized symmetry breaking provides a clear signature that the system undergoes a phase transition at this point.

\section{Phase Transition}\label{part4}

The study of phase transitions in non-Hermitian quantum systems has emerged as a rapidly growing frontier in quantum many-body physics, offering new perspectives on critical phenomena beyond the framework of conventional Hermitian theories. Central to these developments are parity-time ($\mathcal{PT}$) symmetric Hamiltonians, whose phase transitions are characterized by spontaneous breaking of $\mathcal{PT}$ symmetry and the appearance of exceptional points—non-Hermitian spectral singularities at which both eigenvalues and eigenvectors simultaneously coalesce \cite{Bender1998,Bender2007,Ashida2020}. Although the spectral signatures of these transitions, particularly the emergence of complex-conjugate energy pairs, are well established, identifying their manifestations in experimentally accessible observables remains a significant challenge. In large interacting systems, this difficulty is compounded by the exponential growth of the Hilbert-space dimension, which severely restricts the applicability of conventional exact diagonalization methods.

To overcome these limitations, quantum states in neural-networks based on Restricted Boltzmann Machines (RBMs) provide a powerful variational framework for investigating large-scale non-Hermitian many-body systems \cite{Carrasquilla2023}. In this section, we employ the RBM-VMC approach to characterize the non-Hermitian phase transition through longitudinal magnetization $M_z=\frac{1}{N}\sum_{i=1}^{N}\langle \sigma_i^z\rangle$.

As indicated by the ground-state energy spectrum (see Fig\ref{f6}), the system exhibits a symmetry-breaking transition near the critical field $h_c \approx 0.3$. The natural question is whether this transition is also reflected in the magnetic response of the system. Since magnetization constitutes a fundamental order parameter in quantum spin systems, its behavior across the transition provides direct insight into the interplay between quantum fluctuations, non-Hermiticity, and $\mathcal{PT}$-symmetry breaking.

\begin{figure}
    \centering
    \includegraphics[width=1\linewidth]{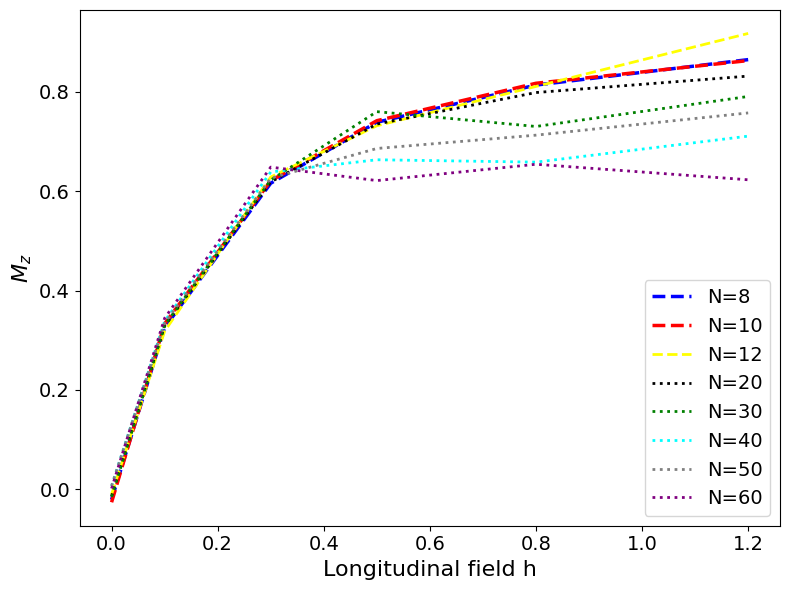}
    \caption{\textbf{Longitudinal magnetization vs.$h$.}  We displays the longitudinal magnetization of the system for different system sizes. We observe that the system exhibit a quantum phase transition around the critical $h=0.3$ as predicted by the ground state energy.
    The fixed parameters for this simulation are $N = 100$ , $\eta = 1.6$, $\xi= 0.16$ }%
    \label{f7}
\end{figure}

To investigate this, we map the ground-state properties of the system over a dense grid of the longitudinal field $h$ [cf.~Fig.~\ref{f7}]. In particular, the site-resolved longitudinal magnetization $M_z(j)=\langle \sigma_j^z \rangle$ is computed to examine how local fields and boundary effects influence the spatial structure of magnetic order. This allows us to resolve how the system develops spatially inhomogeneous magnetization profiles across the transition and how local pinning or suppression of order emerges in different sublattices.

Within this framework, we demonstrate that the RBM variational ansatz faithfully captures the macroscopic signatures associated with the exceptional point. Overall, this approach provides a practical diagnostic tool for identifying $\mathcal{PT}$-symmetry breaking without requiring direct access to the full energy spectrum, while also establishing neural-network quantum states as a powerful framework for probing critical behavior in non-Hermitian quantum many-body systems.

\begin{figure}
    \centering
    \includegraphics[width=1\linewidth]{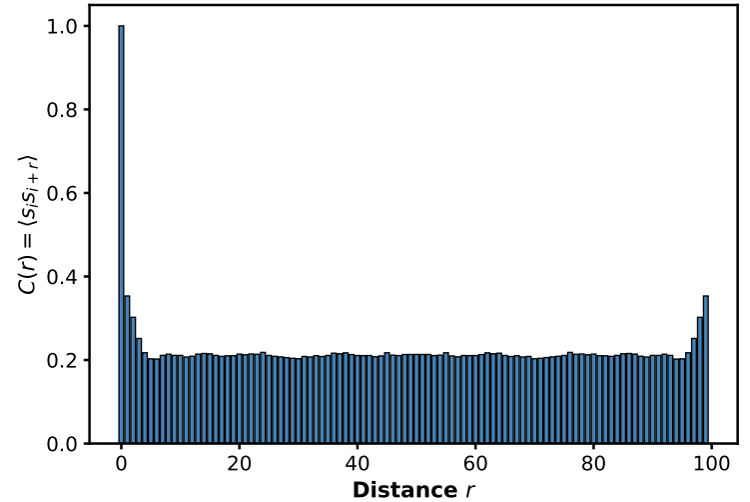}
    \caption{\textbf{spin-spin correlation function.}  We displays the spatial propagation of correlations which serves as a sensitive indicator of criticality and enables a detailed characterization of the quantum phase transition.
    The fixed parameters for this simulation are $N = 100$ , $\eta = 1.6$, $\xi = 0.16$  and $h=0.1$. }%
    \label{f8}
\end{figure}

The structural changes in the many-body wave function become evident through real-space observables. As shown in Fig.~\ref{f8}, spatial correlations are characterized using the spin–spin correlation function,$C_{zz}(r)=\frac{1}{N}\sum_{j}\langle \sigma_j^z \sigma_{j+r}^z \rangle,$ which allows us to quantify the spatial decay of correlations and diagnose the emergence of localized quantum critical regions.

This observable provides direct insight into how correlations propagate across the lattice and how they are modified in the vicinity of the phase transition.

Figure~\ref{f8} presents the equal-time longitudinal spin-spin correlation function,
\begin{equation}
C_{zz}(r)=\frac{1}{N}\sum_{j}\langle \sigma_j^z \sigma_{j+r}^z \rangle,
\end{equation}
computed for a chain of $N=100$ spins with $h=0.1$, $\eta=1.6$, and $\xi=0.16$. This quantity provides direct information on the spatial organization of the many-body ground state reconstructed by the RBM variational ansatz and serves as a sensitive probe of magnetic ordering and quantum criticality.

As shown in Fig.~\ref{f8}, the correlation function reaches its maximum at zero separation, corresponding to the trivial self-correlation, and subsequently decreases with increasing distance before approaching an approximately constant finite value. The persistence of finite correlations at large separations indicates that the ground state preserves long-range ferromagnetic correlations despite the presence of quantum fluctuations generated by the transverse field. This behavior demonstrates that the longitudinal magnetic field stabilizes an ordered magnetic phase in which spin alignment extends throughout the lattice.

The nearly distance-independent plateau observed at large separations further suggests that the correlation length is comparable to the system size, implying that magnetic correlations propagate across the entire chain. Such long-range coherence is fully consistent with the finite longitudinal magnetization reported in Fig.~\ref{f7} and confirms that the RBM variational wave function accurately captures the collective nature of the many-body ground state. The slight enhancement of the correlations at the largest separations originates from the periodic boundary conditions ($\sigma_{N+1}=\sigma_1$), for which spins separated by distances approaching the system size become nearest neighbors again.

From the perspective of non-Hermitian quantum criticality, the spin-spin correlation function provides an independent real-space signature of the underlying PT-symmetry properties. For the present parameter set ($h<h_c$), the absence of a rapid exponential decay or oscillatory behavior indicates that the system remains in the ordered phase preceding the PT-symmetry-breaking transition. As the longitudinal field approaches the critical value ($h_c\approx0.3$), significant modifications of the correlation profile are expected, including the suppression of long-range order and the emergence of enhanced critical fluctuations associated with the exceptional point. Consequently, the behavior of $C_{zz}(r)$ independently corroborates the phase transition inferred from both ground-state energy and longitudinal magnetization, further demonstrating the ability of the RBM framework to characterize quantum critical behavior in large non-Hermitian many-body systems.

\section{Simulation Parameters}

\begin{table}[H]
\centering
\caption{\textbf{Simulation parameters used in the RBM-VMC calculations.}}
\label{tab1}
\renewcommand{\arraystretch}{1.15}
\begin{tabular}{|l|c|}
\hline
\textbf{Parameter} & \textbf{Value} \\
\hline

\multicolumn{2}{|c|}{\textbf{Neural-network architecture}} \\
\hline
Ansatz & $\mathcal{PT}$-symmetric RBM \\
Number of hidden units & 34 \\
Number of hidden layers & 1 \\
Input dimension & 2 \\
Random seed & 123 \\
\hline

\multicolumn{2}{|c|}{\textbf{Variational Monte Carlo}} \\
\hline
Optimizer & Adam \\
Sampling method & Gibbs sampling \\
Learning rate & $10^{-2}$ \\
Monte Carlo samples & 1024 \\
Training steps & $10^{3}$ \\
Regularization parameter ($\alpha$) & 0.2 \\
Log-probability offset & $10^{-15}$ \\
\hline

\multicolumn{2}{|c|}{\textbf{Physical model}} \\
\hline
Exchange coupling & $J=1$ \\
System size & $N=4$--$100$ \\
\hline

\end{tabular}
\end{table}

Unless otherwise stated, all numerical results presented in the following sections were obtained using the simulation parameters listed in Table~\ref{tab1}. For each parameter set, the RBM was independently optimized until convergence of the variational energy was achieved.

\section{Concluding Remarks}\label{part5}

In this work, we successfully implemented a  Restricted Boltzmann Machine variational ansatz combined with Variational Monte Carlo sampling to map out the ground-state properties and critical behavior of a 1D non-Hermitian transverse-field Ising chain subjected to a longitudinal magnetic field. Our network architecture demonstrated high precision, yielding ground-state energies that perfectly match exact diagonalization benchmarks for smaller lattices, and effectively scaling to macroscopic system sizes up to $N=100$.  We demonstrated that the introduction of a longitudinal field $h$ induces a localized phase transition and spontaneous symmetry breaking around $h_c \approx 0.3$, a phenomenon verified through both complex energy scaling and longitudinal magnetization profiles. This approach establishes neural-network quantum states as a powerful, scalable vehicle for probing criticality, tracking $\mathcal{PT}$-symmetry breaking transitions, and exploring exotic quantum phases in non-Hermitian many-body systems.

\bibliography{bibliography}

\end{document}